\documentclass{ws-procs9x6}
\usepackage{pstricks,pst-node,pst-text,pst-3d}
\usepackage{moreverb,epsfig,setspace}
\usepackage{subfigure}
\usepackage{color}

\begin{document} 
\input{col.df}
\definecolor{dgreen}{rgb}{0.1, 0.7, 0.1}
\newcommand{\dgreen}{\color{dgreen}}
\definecolor{lgreen}{rgb}{0.0, 0.5, 0.5}
\newcommand{\lgreen}{\color{dgreen}}

\newcommand{\hspl}{\hspace*{1.5cm}}
\newcommand{\hspm}{\hspace*{0.7cm}}
\newcommand{\hsps}{\hspace*{0.3cm}}
\newcommand{\hspt}{\hspace*{0.1cm}}
\newcommand{\hspz}{\hspace*{0.0cm}}
 \newcommand{\bflr}{\begin{flushright}}
 \newcommand{\eflr}{\end{flushright}}
 \newcommand{\bmini}{\begin{minipage}}
 \newcommand{\emini}{\end{minipage}}
\newcommand\del{\partial}
\newcommand\nn{\nonumber}
\newcommand{\Tr}{{\rm Tr}}
\newcommand{\mat}{\left ( \begin{array}{cc}}
\newcommand{\emat}{\end{array} \right )}
\newcommand{\vect}{\left ( \begin{array}{c}}
\newcommand{\evect}{\end{array} \right )}
 
  \newcommand{\btab}{\begin{tabbing}}
  \newcommand{\etab}{\end{tabbing}}
\newcommand{\bitem}{\begin{itemize}}
\newcommand{\eitem}{\end{itemize}}
\newcommand{\dl}{    $ }
\newcommand{\dr}{ $   }

\newcommand{\be}{  \begin{eqnarray}}
\newcommand{\ee}{\end{eqnarray}  }

\definecolor{Bittersweet}   {cmyk}{0,0.75,1,0.24}
\definecolor{dgreeni}{rgb}{0.2, 0.8, 0.24}

\definecolor{dgreenf}{rgb}{0.7, 0.85, 0.75}

\newcmykcolor{colbsc}{0 0.75 1 0.24}
\newrgbcolor{colgi}{0.1 0.2  0.125}
\newrgbcolor{dgreen}{0.1 0.6 0.1}

\newcommand{\colbo}{\color{Bittersweet}}
\newcommand{\colbs}{\color{Bittersweet}}
\newcommand{\colgr}{\color{Blue}}
\newcommand{\colyg}{\color{DarkOrchid}}
 \newcommand{\colgy}{\darkblue}
\newcommand{\colrd}{\darkblue}

 \title{LESSONS FROM RANDOM MATRIX THEORY FOR \\  QCD AT FINITE DENSITY}

\author{K. Splittorff }

\address{The Niels Bohr Institute,\\
Blegdamsvej 17, Dk-2100, Copenhagen, Denmark\\
E-mail: split@nbi.dk}

\author{J.J.M. Verbaarschot}

\address{Stony Brook University,\\
Stony Brook, NY 11749-3800, USA\\
E-mail: jacobus.verbaarschot@stonybrook.edu}

\begin{abstract} 
In this lecture we discuss various aspects  of QCD at nonzero chemical
potential, including its phase diagram and the Dirac spectrum,  and 
summarize what chiral random matrix theory has contributed to this subject.
To illustrate the importance of the phase of the fermion
determinant, we particularly highlight the differences between 
QCD and phase quenched QCD.  
\end{abstract}

\bodymatter
\section{Introduction}
QCD at nonzero baryon chemical potential has turned out to be
 particularly challenging,
and as of today, the phase diagram of QCD
in the chemical potential temperature plane is far from being understood. 
There is not even agreement
of its gross features such as the existence of a critical endpoint
\cite{misha-lat,fodor,philip}. 
The reason for these difficulties is that the Dirac operator at nonzero
chemical potential is nonhermitean resulting in a complex determinant 
so that the partition function cannot be evaluated by Monte-Carlo simulations.
In this lecture we compare the QCD partition function to the QCD like
partition function that only differs by the absence of the phase of the
fermion determinant. This theory, known as phase quenched QCD, can be
studied by means of lattice QCD simulations 
and has a phase diagram that is certainly completely different from QCD.
This makes it clear that the phase
of the fermion determinant is absolutely essential for the physics
of QCD at nonzero chemical potential. 

In the second half of this
lecture we will discuss various aspects of the phase of the fermion
determinant and analyze its connections with the spectrum of the Dirac operator.
We will do this in the context of chiral random matrix theory \cite{V,SV}
at nonzero chemical
potential \cite{misha,hov,O}. 
In the microscopic domain \cite{SV}, this theory is equivalent to QCD and, 
in spite of the
fact that it can be solved analytically, it shares the sign problem and
the physical effects of the phase of the fermion determinant with QCD.

\section{Phase Diagram of QCD}

The phase diagram of QCD in the temperature-chemical potential plane
can only be obtained by means of nonperturbative methods. The method
that has had the most impact is lattice QCD, but because it
requires a real action, only the temperature
axis has been studied extensively. The consensus is 
that, in the temperature range of  150-200 MeV, 
QCD undergoes a crossover transition from a chirally
broken phase to a chirally symmetric phase. The picture at nonzero
chemical potential is much less clear. Although significant progress has
been made for small chemical potentials, questions about the structure
of the phase diagram along the chemical potential axis could not be
answered based on first principles  (except for asymptotically large chemical
potentials \cite{son}). 
That is why most of our knowledge of QCD at nonzero baryon
density is based on models and general properties of the phases of QCD.

\begin{figure}
\centerline{\includegraphics[width=14cm]{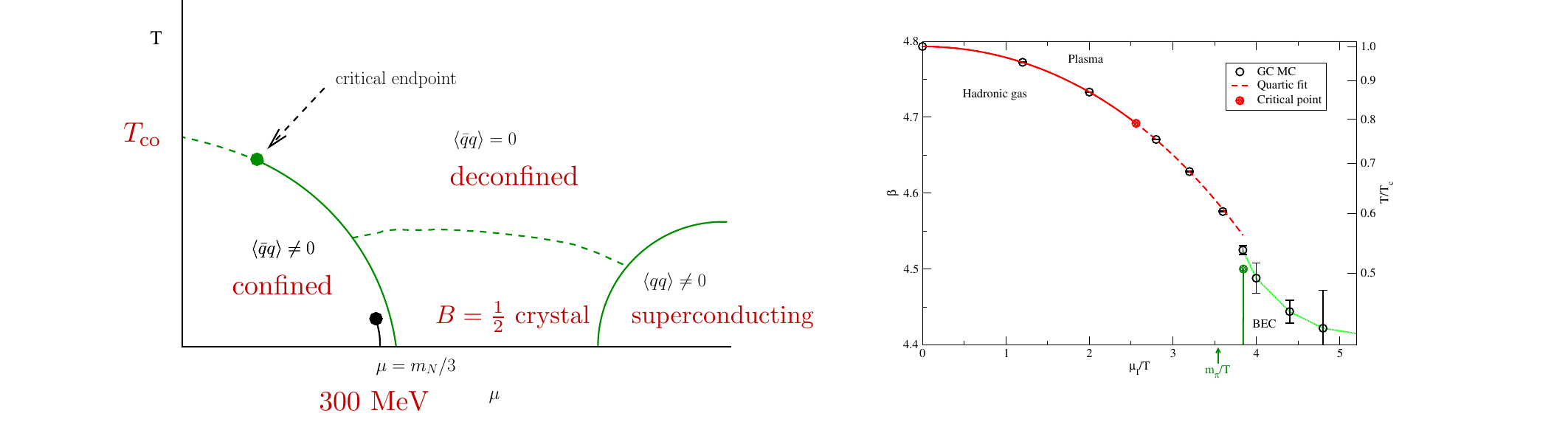}}
\caption{The phase diagram of QCD in the temperature baryon chemical potential
plane (left) and in the temperature isospin chemical potential plane 
(right). The data points in the right figure are from lattice QCD simulations
\cite{wenger}.}
\label{fig1}
\end{figure}

\begin{figure}
\includegraphics[width=5.5cm,height=5cm]{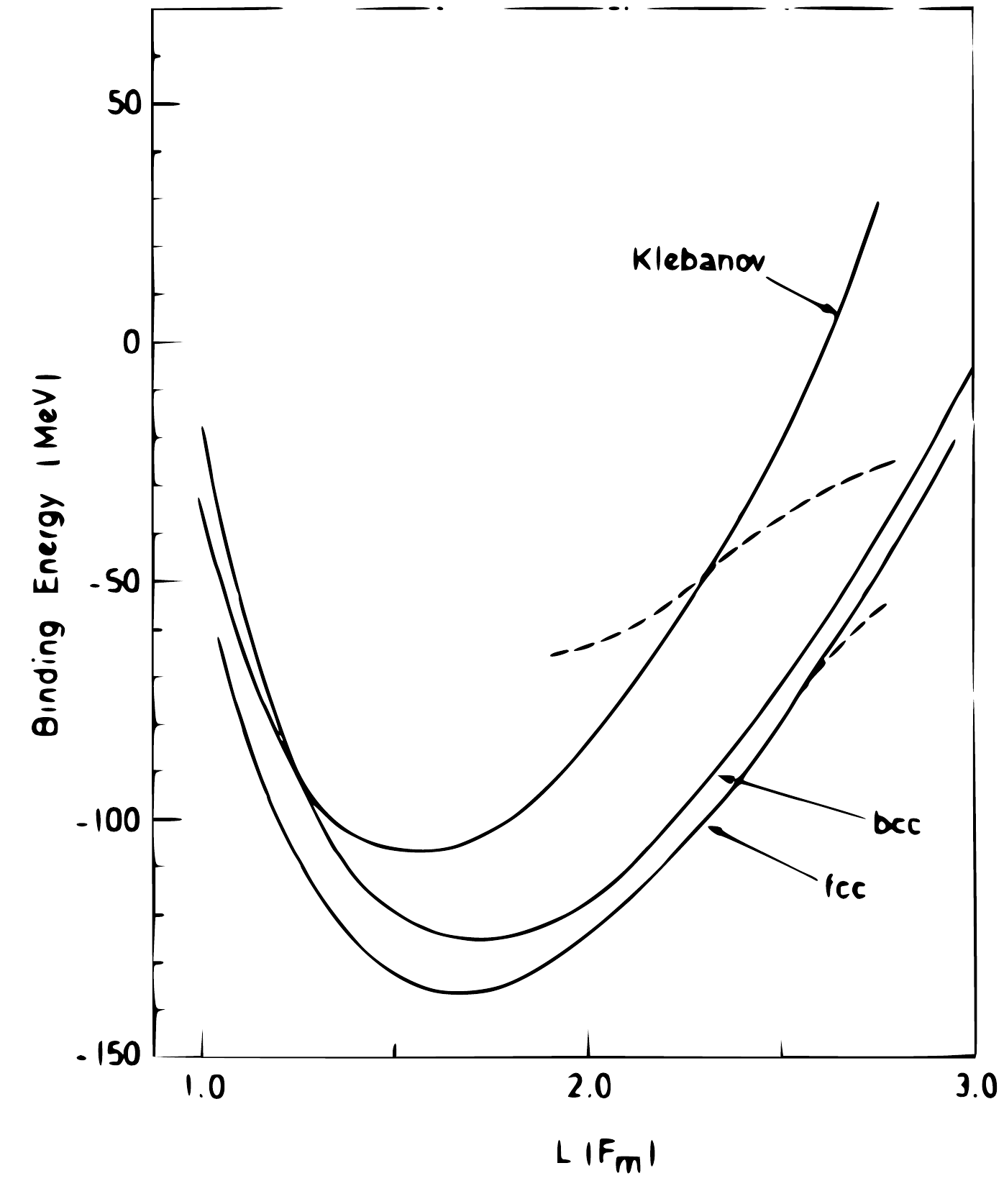}
\includegraphics[width=5.5cm,height=5cm]{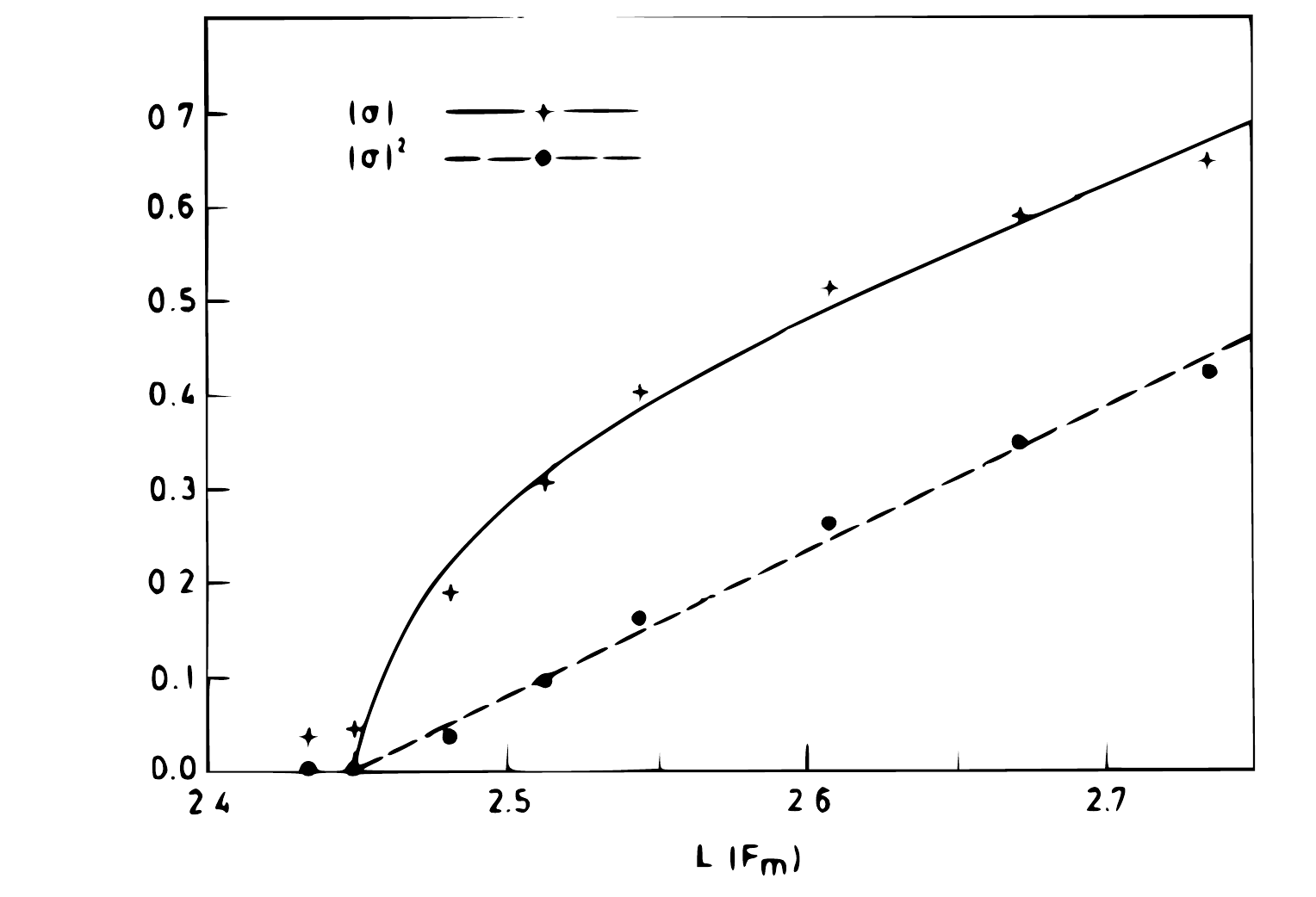}
 \caption{Energy of a unit cell of a Skyrmion crystal (left) for a cubic
crystal with cubic boundary conditions 
(obtained in \cite{klebanov}) and  for a bcc and fcc crystal (obtained in
\cite{jackson}) versus the size of the unit cell. In the right figure
we show the average value of the $\sigma$-field for an fcc crystal 
\cite{jackson}. }
\label{fig2}
\end{figure}

A tentative phase diagram of QCD in 
the chemical potential temperature plane is shown in Fig. \ref{fig1} (left). 
Among others
we conjecture a chirally symmetric crystalline phase made out of
$B=\frac 12 $ objects \cite{goldhaber}. The best evidence for  the 
existence of such phase is based on the Skyrme model where this phase
exists as a strongly bound crystalline state with a binding energy of 135 MeV 
per nucleon (see Fig. \ref{fig2} (left)). 
The chiral restoration phase transition to this phase is of second order 
for a face
centered cubic crystal \cite{jackson,jackson1} but is of first order for a 
cubic crystal \cite{klebanov}. The restoration of
chiral symmetry in the dense phase is illustrated by the 
 vanishing of the average  $\sigma$-field over a unit cell (see Fig. 
\ref{fig2} (right)).
 When the temperature is increased, we expect
that this crystal will melt at a temperature that is of the order of the 
binding  energy. 
This phase has also been identified as the quarkyonic phase
\cite{Mclerran1,Mclerran2}.

Our aim is to understand QCD at $\mu \ne 0$ from
first principles, i.e. starting from the partition function
\be
Z_{{\rm QCD}} = \langle \prod_f \det(D + m_f + \mu \gamma_0) \rangle,  
\ee
but the nonhermiticity of \dl D+ \mu \gamma_0 \dr makes this a very 
challenging problem indeed. 
Let us first discuss the effect of the phase of the fermion determinant
by studying the theory where this phase has been quenched.
For $N_f=2$ this partition function, known as phase quenched QCD, is given by 
\be
Z_{|{\rm QCD}|} &=& \langle |\det(D + m + \mu \gamma_0)|^2 \rangle 
= \langle \det(D + m + \mu \gamma_0)\det(D + m  -\mu \gamma_0) \rangle 
\hsps\nn\\
\ee
Therefore, the chemical potential of phase quenched QCD can be interpreted as
an isospin chemical potential \cite{alford}, and at low enough temperatures,
pions will Bose condense for \dl \mu > m_\pi/2 \dr \cite{kstvz,TV,SS}. 
This partition function can be simulated by Monte-Carlo methods 
\cite{hands,kogut1,kogut2,kogut3,wenger}
(see Fig. \ref{fig1}) and its
phases are in agreement with theoretical expectations 
based on a mean field treatment of a chiral Lagrangian \cite{kstvz,STV2,DN}.

We conclude that for \dl \mu > m_\pi/2\dr the
phase factor quark determinant completely changes the phase diagram.  
The would be pion condensate is
nullified by this phase factor after averaging.
To better understand the effect of nonhermiticity, we will analyze the 
QCD partition function in the microscopic domain where it is equivalent
to a chiral random matrix theory. This theory is
analytically solvable in the nonperturbative
domain of QCD, but the sign problem has all complications of QCD.

\section{Random Matrix Model at $\mu \ne 0$}
 
A chiral Random Matrix Theory (chRMT) is obtained by replacing the matrix
elements of the Dirac operator by Gaussian random numbers \cite{V,SV}
\be
D= \mat 0 & iW^\dagger + \mu^\dagger \\ iW +\mu & 0 \emat
\ee
with \dl W \dr a complex \dl n\times(n+\nu)\dr matrix, and \dl \mu \dr is
a multiple of the identity \cite{misha}
or an arbitrary random matrix  \cite{O}.
The random matrix partition function
in the sector of topological charge $\nu$ is given by
\be
Z_\nu(m_f; \mu) = \langle \prod_f \det (D +m_f) \rangle ,
\vspace*{-0.3cm}
\ee
where the average is over the Gaussian distribution of $W$.
\noindent
This partition function has the global symmetries and transformation
properties of QCD. At fixed \dl \theta \dr-angle it is given by
\be
Z(m_f,\theta; \mu) = \sum_\nu {\cal N}_\nu(\mu) e^{i\nu\theta} Z_\nu(m_f; \mu)
\ee
with ${\cal N}_\nu(\mu)$ s normalization constant that may depend
on $\mu$ and $\nu$.

In the microscopic domain of QCD, defined by 
\be
m_\pi^2 \ll \frac 1{\sqrt V}, \qquad \mu^2 \ll \frac 1{\sqrt V},
\qquad V \Lambda_{\rm QCD}^4 \gg 1 
\ee
the mass and the chemical potential dependence of the QCD partition
function is given by the random matrix partition function
\be
Z^{\rm QCD}_\nu(m;\mu) \sim Z^{\rm chRMT}_\nu(m; \mu).\nn
\ee
The reason for this equivalence is that
in this domain, because the Compton wave length of the Goldstone modes
is much larger than the size of the box,
both theories reduce to  the same static chiral Lagrangian.
The Goldstone mass may either refer to physical quarks masses or to 
complex parameters (ghost quark masses) to probe the Dirac spectrum.

Since the kinetic term of  the chiral Lagrangian does not contribute
in a mean field analysis, chRMT correctly describes mean field results
for the low-energy limit of QCD beyond the microscopic domain provided that
$m \ll \Lambda_{\rm QCD}$ and $\mu \ll \Lambda_{\rm QCD}$.

\section{Lessons from Random Matrix Theory}
\subsection{Lesson 1: Homogeneity of Dirac Eigenvalues}

At nonzero chemical potential, the Dirac operator is nonhermitean
with eigenvalues that are scattered in the complex plane. On inspection of
quenched lattice QCD Dirac spectra (see Fig.~3) two properties stand out:
\bmini{3cm}
\includegraphics[width=3.4cm]{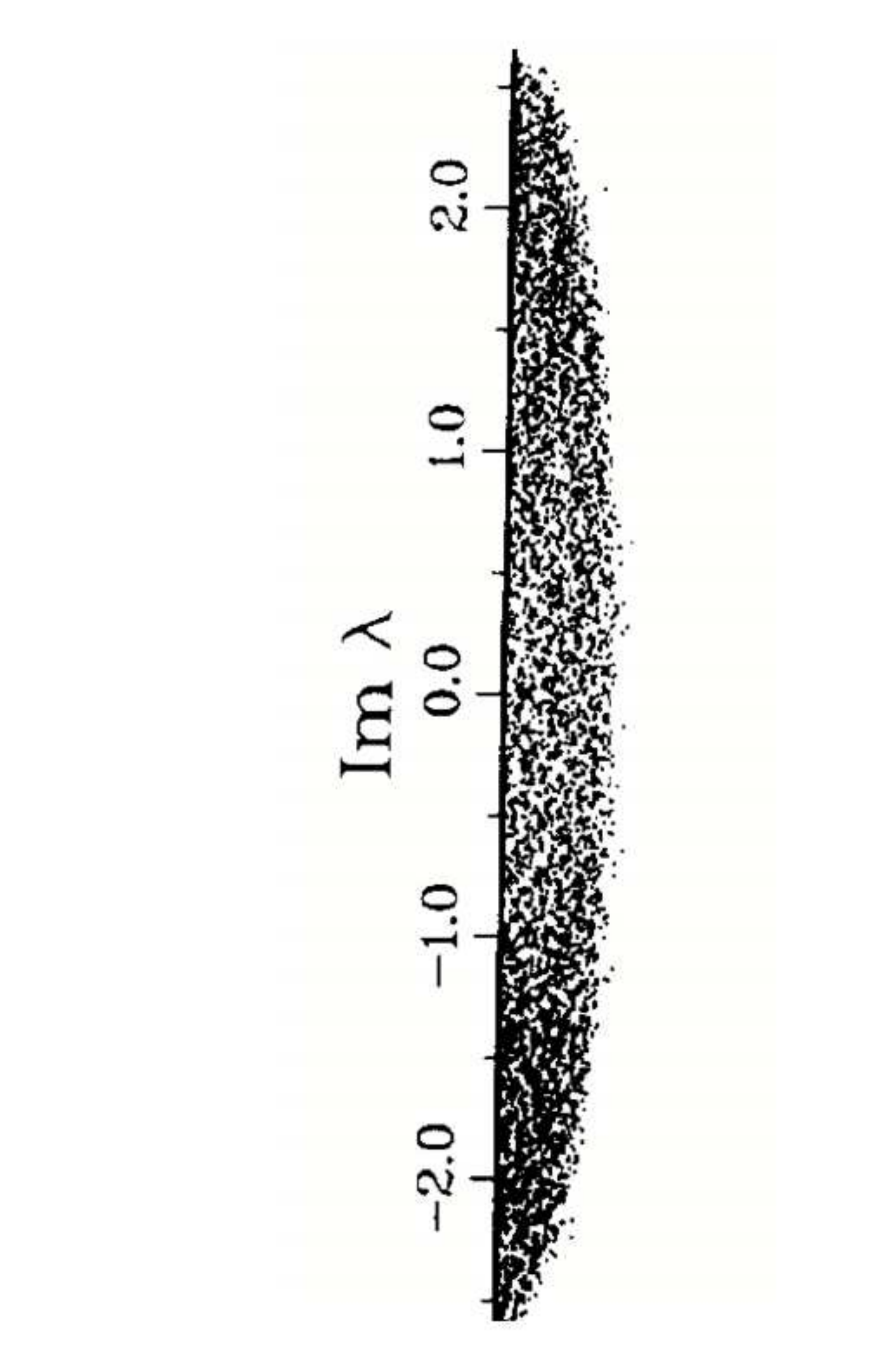}
\linespread{0.707}
\selectfont
{\fontsize{8}{8pt} \selectfont 
Fig. 3. Quenched Dirac eigenvalues on a $ 4^3\times 8 $ 
lattice \cite{all}.}
\vspace*{0.2cm}
\emini\hspace*{0.5cm}
\vspace*{0.2cm}
\bmini{7.9cm}
\vspace*{0.1cm}
i) The Dirac spectrum has a sharp edge, and ii) the distribution of
the eigenvalues is more or less homogeneous. Both properties can
be understood in terms of random matrix theory \cite{misha,hov}. In essence 
they follow from the fact that eigenvalues of nonhermitean random matrices
behave as repulsive  electric charges in the  plane. Because the chiral
condensate is given by
\be
\Sigma(m) = \frac 1V \sum_{\lambda_k} \frac 1{m+i\lambda_k},\nn
\ee
it can be interpreted as the electric field at \dl m\dr of charges
at \dl \lambda_k\dr. Therefore, a direct consequence of the homogeneity of the
Dirac spectrum is that the quenched chiral
condensate increases linearly with \dl m \dr when $m$ is inside the domain
of the eigenvalues.
\emini

\subsection{Lesson 2: Quenched Limit}

The quenched limit is the limit where the fermion determinant is
ignored in generating the statistical ensemble. For zero chemical potential
this has been a reasonable approximation, but at nonzero chemical 
potential, the limit of no fermion determinant is {\colbs not} given by\\
$
\hspace*{3cm} \lim_{n\to 0} \langle (\det(D+\mu\gamma_0 +m))^n\rangle,
$\\ 
but rather by\\
$
\hspace*{3cm}\lim_{n\to 0} \langle |\det(D+\mu\gamma_0 +m)|^n\rangle,
$\\
i.e. the quenched limit of phase quenched QCD \cite{misha}.
Although similar ideas were proposed before \cite{gibbs,gocksch},
this result was first demonstrated convincingly for random matrix theory
where the averages can be evaluated analytically.

Because phase quenched QCD is QCD at nonzero isospin chemical potential,
 a phase transition to a pion condensate
occurs at $\mu = m_\pi/2$ for low temperatures. 
This phase is absent in full QCD
where, at low temperatures a phase transition to a phase 
with nonzero baryon density 
occurs at 
$\mu = m _N/3$. We conclude that quenching completely fails for
$\mu \ne 0$.
\vspace*{-0.3cm}

\subsection{Lesson 3: Width of the Dirac Spectrum}

The phase quenched partition function undergoes a phase transition 
at \dl\mu = m_\pi/2 \dr. In terms of Dirac eigenvalues the only 
critical point occurs
\bmini{5.3cm}
\vspace*{-0.1cm}
\centerline{\includegraphics[width=5.0cm]{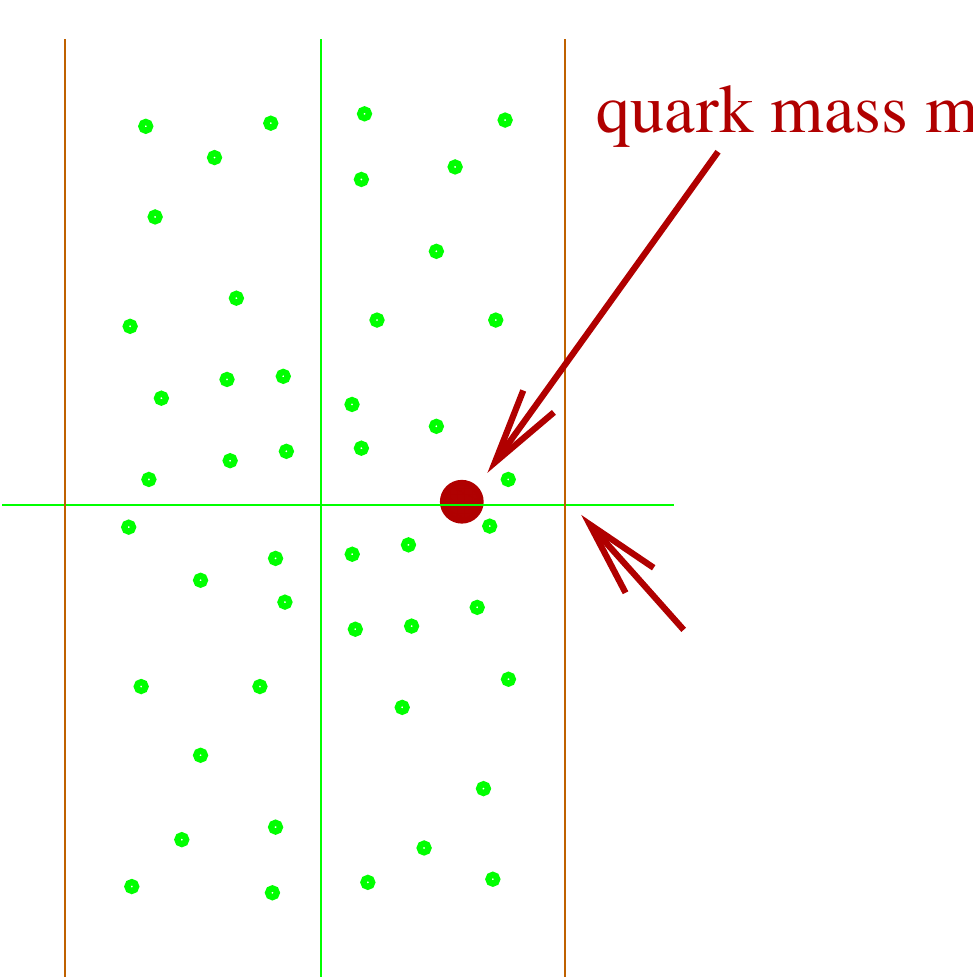}\hspace*{0.5cm}}
\linespread{0.707}
\selectfont
{\fontsize{8}{8pt} \selectfont
\noindent
Fig. 4. Scatter plot of Dirac eigenvalues for quenched or phase quenched
QCD at nonzero chemical potential.}
\emini\hspace*{0.5cm}
\rput[l](-2.5,-0.3){ \dl \frac{\mu^2 F^2}{2\Sigma}\dr}
\bmini{5.6cm}\vspace*{0.1cm}
when the quark mass hits the domain of eigenvalues
(remember that it has a sharp edge). Therefore these two points have
to coincide \cite{gibbs}.
This leads to the following condition for the half-width $m_c$ of the
Dirac spectrum:\vspace*{-0.2cm}
\be
m_\pi^2 = \frac {2m_c \Sigma}{F^2} = 4\mu^2 .
\ee
This result has also been derived from the static limit of the chiral
Lagrangian for phase quenched QCD \cite{TV} which contains 
two low-energy constants, $\Sigma$ and $F_\pi$, as parameters. 
Since, at nonzero chemical
potential 
\vspace*{-0.25cm}
\emini\\
the low-lying Dirac spectrum is characterized by two quantities:
the density of eigenvalues and the width of the spectrum, it is possible
to extract $\Sigma$ and $F_\pi$ from these variables.
More sophisticated methods to extract these constants from the low lying
Dirac spectrum have been proposed and have been successfully applied to
lattice QCD \cite{tilo,tilo-james,nbi1,nbi2,nbi3,nbi4,nbi5,nbi6,nbi7}.
For sufficiently small  chemical
potential, when perturbation theory
applies, the width of the Dirac spectrum increases 
linearly with \dl \mu \dr rather than quadratically.
\vspace*{-0.1cm}
\subsection{Lesson 4: Infrared Dominance}

The question we wish to address in this section is whether the fluctuations
of the fermion determinant given by
\be
\det (D+m +\mu\gamma_0) = \prod_k(\lambda_k +m)
\ee
are dominated by the infrared part of the Dirac spectrum. At $\mu = 0$,
because the low-energy limit of QCD is given by chiral perturbation
theory for sufficiently small quark masses, this is apparently the case.
This is less clear at $\mu \ne 0$ when the eigenvalues are complex. 
In particular, because the eigenvalue density for large $\lambda$ increases 
as $ \sim V \lambda^3 $ one would expect that also the ultraviolet part
of the Dirac spectrum contributes to the phase of the determinant. Although 
a nonzero chemical potential does not introduce new infinities once the
theory is regularized at $\mu =0$, one cannot exclude finite contributions
from large Dirac eigenvalues.

When the chemical potential and the quark masses
are in the microscopic domain of QCD, though, we can use random  matrix theory
to show that the average determinant
is determined by the small Dirac eigenvalues. Further evidence for
the infrared dominance of this quantity comes from lattice QCD at
imaginary chemical potential, where
a small number of low-lying eigenvalues reproduce the
random matrix result \cite{svet}. 

Because of agreement with lattice QCD we know that chiral perturbation
theory can be applied to phase quenched QCD when both the chemical potential
and $m_\pi$ are of order of $F_\pi$ or less. This shows that the magnitude of
the quark determinant is infrared dominated in this domain. 
Since the full QCD partition function in this domain does not depend
on the chemical potential at low temperatures, in agreement with 
chiral perturbation theory, we conclude that the $\mu$-dependence of
the the phase of the fermion determinant resides in the infrared part
of the Dirac spectrum.

\subsection{Lesson 5: Failure of Banks-Casher at \dl \mu \ne0 \dr}

The Banks-Casher relation \cite{BC} states that
\be
\Sigma = \lim_{m\to 0} \lim_{V\to \infty}\frac {\pi \rho(m)}V,
\ee
where $\rho(\lambda)$ is the density of Dirac eigenvalues and $V$ is the 
volume of space-time. Although originally intended for a Hermitian Dirac
operator, this relation correctly gives a vanishing chiral condensate 
for phase quenched QCD at $\mu \ne 0$.
However, for full QCD at \dl \mu \ne 0 \dr, the chiral condensate has
a discontinuity when the quark mass crosses the imaginary axis, but 
it does so without the occurrence of an accumulation of eigenvalues. The
alternative mechanism that is at work has been understood in 
detail in random matrix theory \cite{OSV,OSV-poly}. For a comprehensive review
we refer to the  talk by Splittorff also in this volume \cite{splittalk}.
Below we will illustrate this mechanism for QCD in one dimension
which can also be viewed as a random matrix model. 

\begin{picture}(400,150)(10,10)
\put(10,10)
{\includegraphics[height=5.2cm]{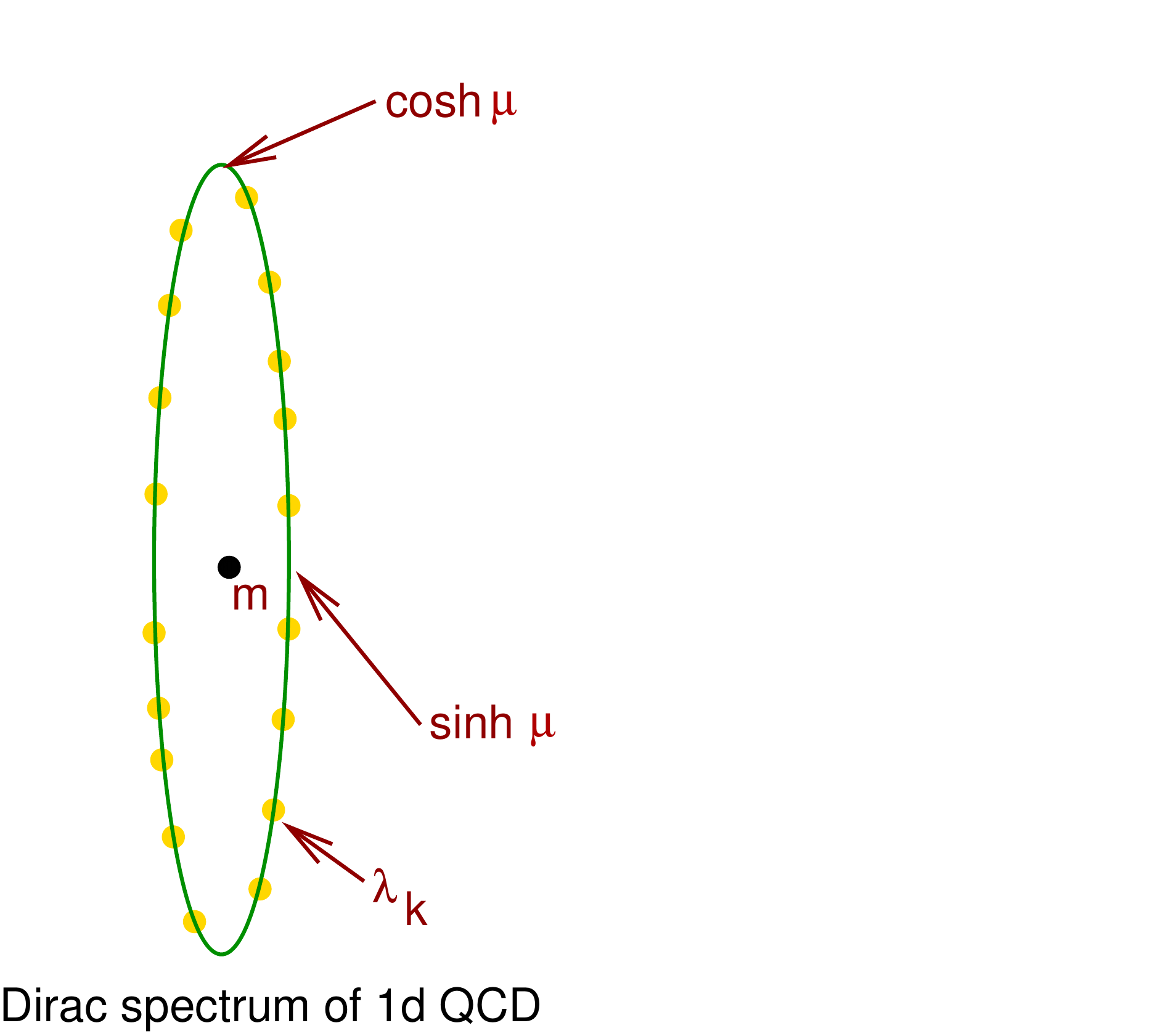}}\hspace*{-1cm}
\put(200,150){\includegraphics[height=4.2cm,angle=-90]{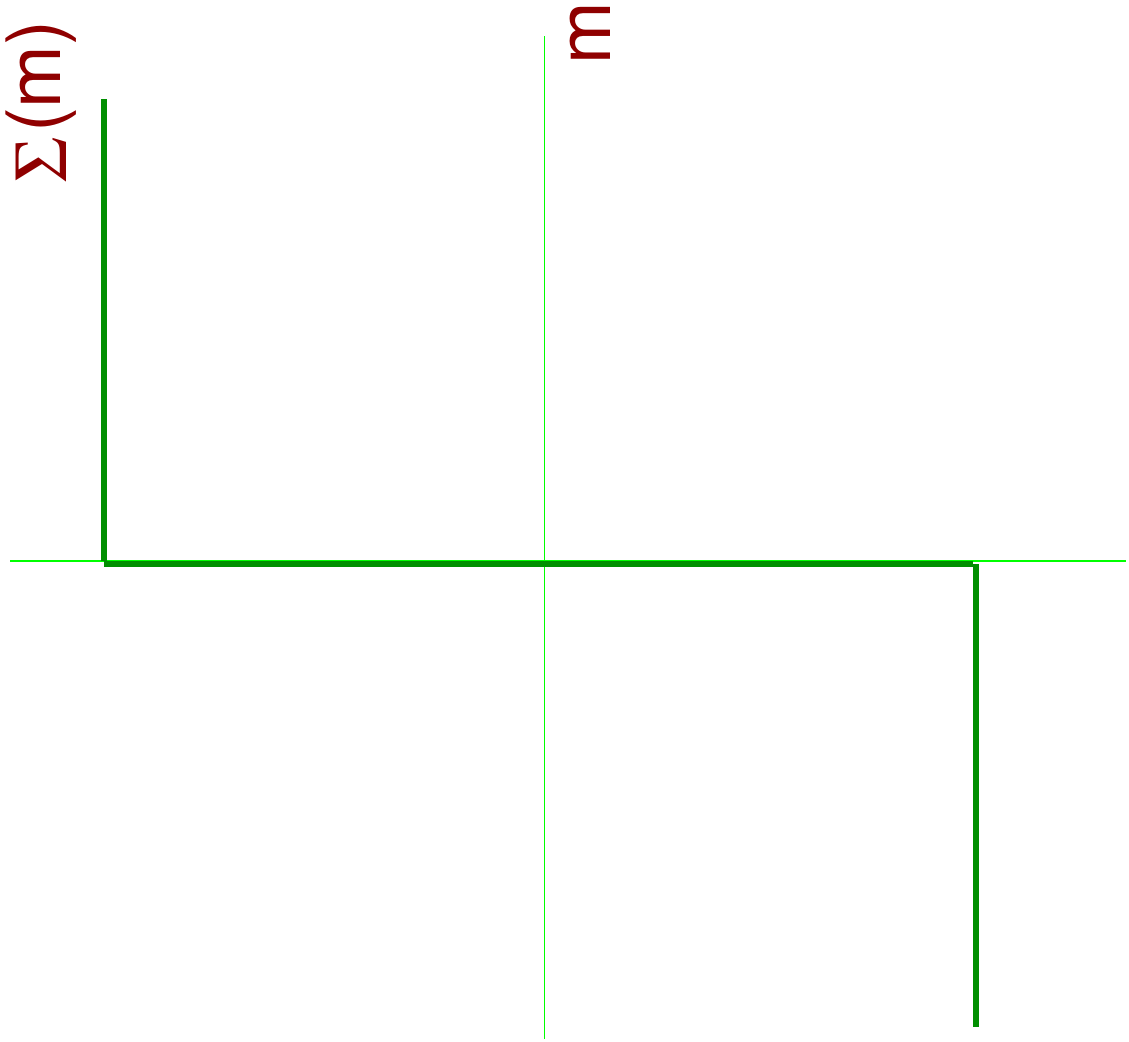}}
\end{picture}
\linespread{0.707}
\selectfont
{\fontsize{8}{8pt} \selectfont
\noindent        Fig. 5.
Schematic plot of eigenvalues of the Dirac operator for 
lattice QCD in one dimension (left). The yellow dots denote the position of
the eigenvalues for a single gauge field configuration, whereas the green
ellipse shows the support of the spectrum in the thermodynamic limit.
The chiral condensate for one flavor versus the quark  mass is 
shown in the right figure. 
}

\linespread{1}\selectfont
The partition function of lattice QCD in one dimension is given by
\be
Z= \int_{U \in U(N_c)} dU \det D,
\ee
where the integral is over the Haar-measure of $U(N_c)$. The Dirac operator
is given by the $N\times N_c$ matrix (for $N$ lattice points)
 \be
D = \left ( \begin{array}{cccc} 
mI & e^{\mu}      & {\ldots} &  e^{-\mu}U^{\dagger}\\
-e^{-\mu} & mI & \cdots  &0\\ 
\vdots& &&\vdots\\
 0& \cdots&mI&e^\mu\\
-e^{\mu}\rnode{a}{}U/2& \cdots &-e^{-\mu}&mI\\
\end{array} \right).
\ee
The chiral condensate for one flavor is defined by
\be\Sigma(m)  
=\frac {\left \langle \frac 1N\sum_k \frac 1{\lambda_k+m} 
{\prod_k(\lambda_k+m)}\right \rangle }
 {\left \langle
\Rnode{a1}{\prod_k(\lambda_k+m)}\right \rangle }.
\ee
Notice that the determinant has a complex phase. For $U(1)$ the chiral 
condensate can be evaluated analytically \cite{ravagli} with the result
that is shown in the right figure of Fig.~5. The amazing phenomenon,
also known as the ``Silver Blaze Problem'' \cite{cohen},
is that the chiral condensate is continuous when $m$ crosses the ellipse
of eigenvalues, but shows a discontinuity at $m =0$ where there are no
eigenvalues. This can happen because the chiral condensate is determined
by exponentially large (in the number of lattice points) contributions which
cancel to give a finite result for the chiral condensate for
$N \to \infty$. 

\subsection{Lesson 6: Quark Mass and Average Phase Factor}

The severity of the sign problem can be measured through the expectation
value of the average phase factor.
A  physical interpretation is 
obtained by defining the phase with respect to the phase quenched 
partition function:
\be
 \langle e^{2i\theta} \rangle_{\rm pq} 
= \frac{ \langle {\det}^2(D+m +\mu\gamma_0)\rangle}
{ \langle |\det (D+m +\mu\gamma_0)|^2 \rangle}
\equiv\frac{ Z_{\rm QCD}}{Z_{\rm |QCD|}} . \nn
\ee
At nonzero temperature the free energies of QCD and $|{\rm QCD}|$ are different
so that the average phase factor vanishes in the thermodynamic limit.
 
At zero temperature, the free energy of QCD and phase quenched QCD are
the same for $\mu <m_\pi/2$. 
Therefore, in the thermodynamic
limit,  the average phase factor is
one for $\mu < m_\pi/2$.  
At finite volume, a nontrival  
\bmini{6cm}\hspace*{-0.5cm}
  \includegraphics[width=5.0cm]{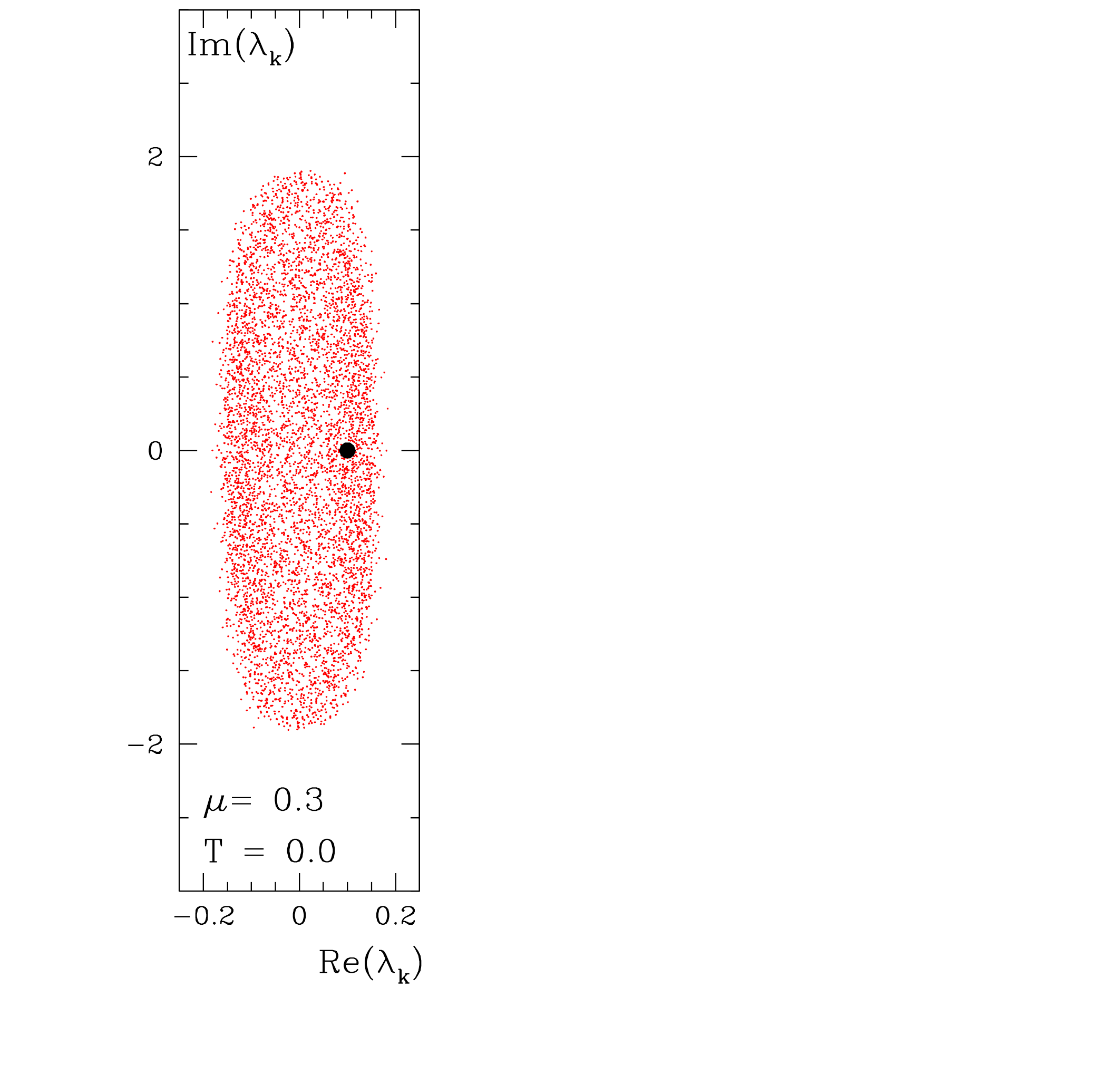}\hspace*{-3.3cm}
  \includegraphics[width=4.8cm]{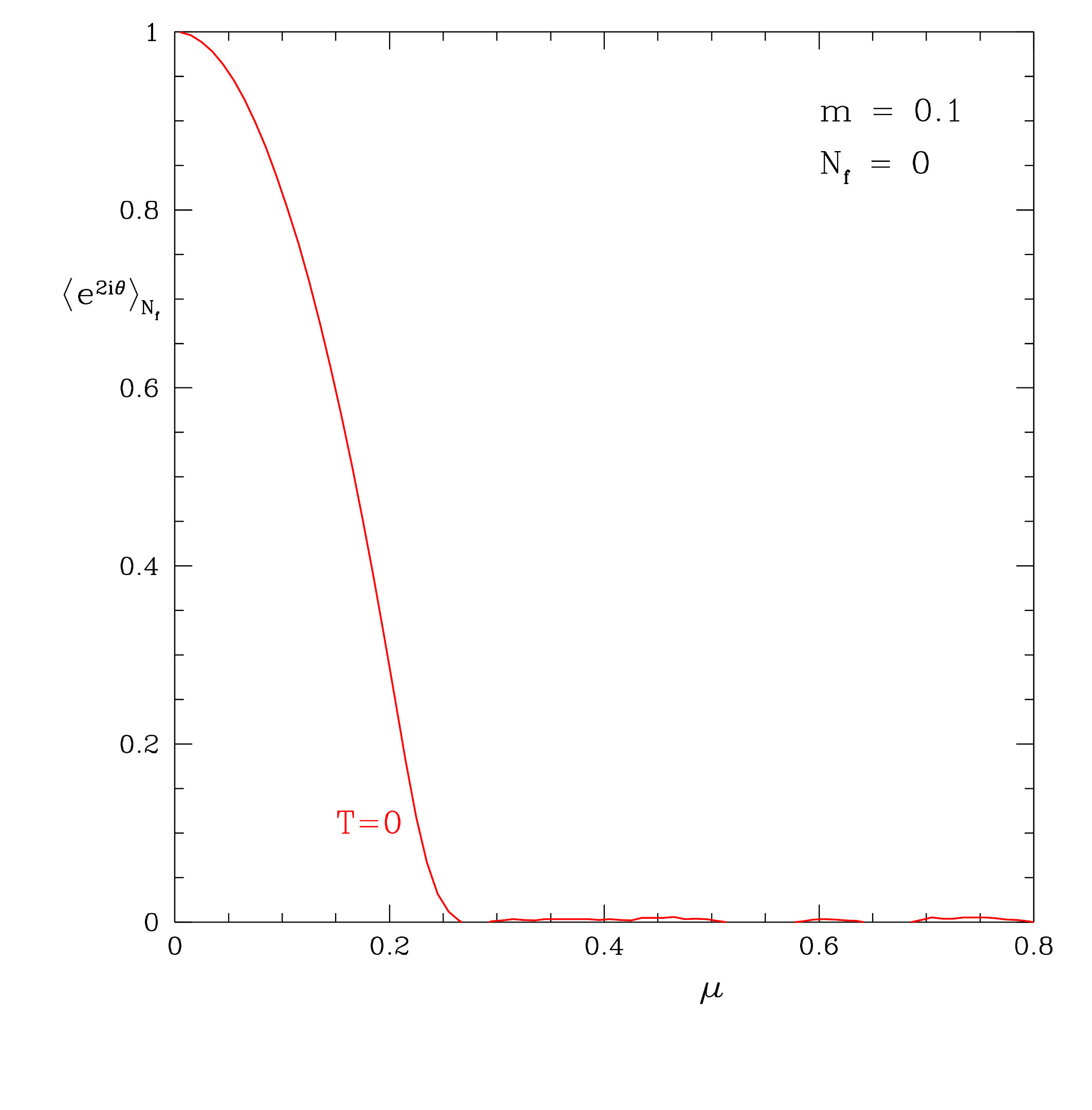}\\
\linespread{0.707}
\selectfont
{\fontsize{8}{8pt} \selectfont
Fig. 6. Scatter plot of Dirac eigenvalues for chiral random matrix
theory at nonzero chemical potential (left) and the average phase
factor as a function of the chemical potential (right) for $m=0.1$.
In the left figure the mass is indicated by a black dot.}
\label{fig6}
\emini
\hspace*{0.5cm}
\bmini{4.9cm}
\vspace*{0.1cm}
$\mu$-dependence
is obtained from the one-loop
corrections which give rise to an $O(1)$ contribution.

In the microscopic limit the average phase is given by 
the random matrix result that  can be evaluated
analytically (see Fig.~6 (right)). 
In the left part of the same figure we
show a scatter plot of the Dirac eigenvalues of 40 $200\times 200$
random matrices. We conclude that, in the microscopic domain of QCD, 
the average phase factor vanishes if
the quark mass is inside the support of the Dirac spectrum.
\emini\vspace*{0.1cm}
 
The average phase factor has been evaluated to one-loop order in chiral perturbation 
theory \cite{splitphase},
and the results are in good agreement with lattice QCD \cite{Allton}.
The mean field limit of this result agrees with 
random matrix theory  in the limit where 
the microscopic variables are large.  
Not surprisingly, also in the domain of chiral perturbation theory, the 
sign problem is most severe when the quark mass is inside the domain of
eigenvalues.

\subsection{Lesson 7: Distribution of Small Dirac Eigenvalues}
One of the greatest successes of chiral random matrix theory has
been the exact description of the distribution of the low-lying 
eigenvalues of the Dirac operator \cite{V,SV,VZ} both at zero 
(see review \cite{wettig}) and at nonzero chemical potential
(see review \cite{akrev}). In this section we show recent
results for the Dirac spectrum at nonzero chemical potential
and topology. Such lattice calculations have become
possible  because of the introduction of an overlap Dirac
operator at nonzero chemical potential \cite{bloch-wettig}. In 
Fig.~7, the radial distribution of the lowest Dirac
eigenvalue is shown for different values of the topological charge in
comparison with analytical results derived in \cite{shifrin}. 

\vspace*{0.3cm}
\centerline{ \includegraphics[width=7cm]{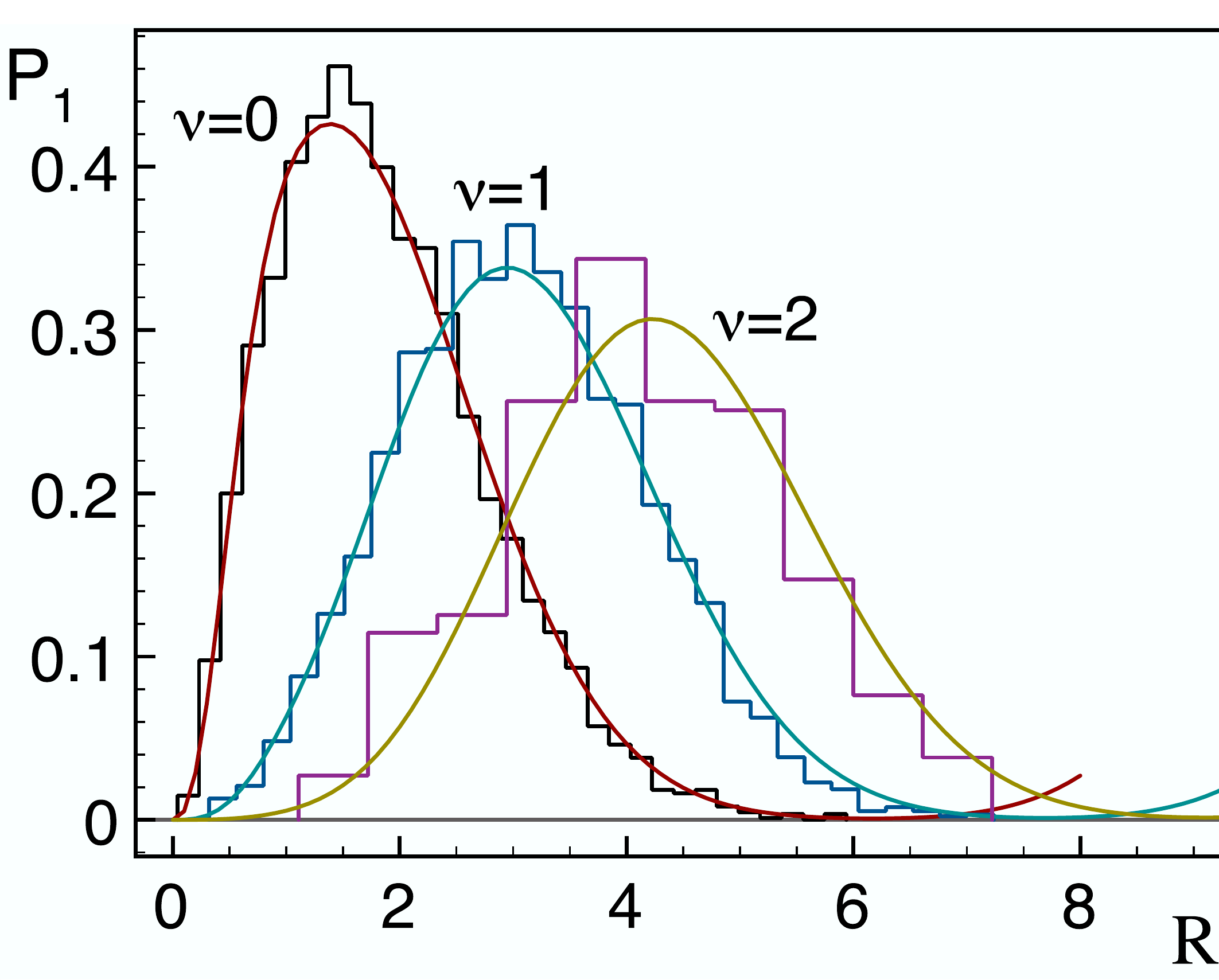}}
\noindent
\linespread{0.707}
\selectfont
{\fontsize{8}{8pt} \selectfont
\noindent
Fig. 7. Radial distribution of smallest Dirac eigenvalue 
for $\mu\neq0$ in different topological charge sectors using a Dirac operator
that satisfies the Ginsparg-Wilson relation (histograms)\cite{shifrin} 
compared to the prediction of chiral random matrix theory (smooth curve).}
 
\subsection{Lesson 8: Equality Two Condensates}

\linespread{1}
\selectfont
The chiral condensate can be calculated in two ways
\be 
\Sigma^{(1)} = \lim_{m\to 0}\lim_{V\to\infty}\frac 1V \left \langle \sum_k 
\frac 1{\lambda_k+m} \right \rangle, 
\Sigma^{(2)}=\lim_{V\to\infty}\lim_{m\to 0}\frac 1V \left \langle \sum_k 
\frac 1{\lambda_k+m} \right \rangle ,\nn \\
\ee
where $\Sigma^{(1)} $ is nonzero because of spontaneous
symmetry breaking, and its value does not depend on the total topological
charge. The second chiral condensate can be expressed as
\be
\Sigma^{(2)} =\lim_{V\to\infty}\frac 1V \frac {\langle \prod_{\lambda_k \ne 0}\lambda_k
\rangle_{\nu =1}} {\langle \prod_{\lambda_k }\lambda_k
\rangle_{\nu =0}} .
\label{ratnu}
\ee 
The reason is that in the sector of topological charge $\nu$, there are $\nu$ 
exact zero modes. 
The equality $ \Sigma^{(1)}= \Sigma^{(2)}$ requires a subtle
reshuffling of the eigenvalues: For $\nu = 1 $ the eigenvalues are shifted
by approximately half a level spacing w.r.t. $ \nu = 0 $  in order to 
satisfy the chiral Ward identity. Therefore, $\lambda_k^{\nu=1} \approx
\sqrt{\lambda_k^{\nu=0}\lambda_{k+1}^{\nu=0}}$.
To correctly normalize the ratio in (\ref{ratnu}) we evaluate it for
a finite Dirac operator with an $N\times (N+\nu) $ nonzero off-diagonal block.
Such Dirac matrix has $\nu$ exact zero eigenvalues (and perhaps additional
paired zero modes which, being a set of measure zero, can be safely ignored).
As it stands, the ratio (\ref{ratnu}) is dimensionally
incorrect. A dimensionally correct ratio is obtained by replacing the
largest squared eigenvalue pair for  $\nu = 0$ by its square root,
$[\lambda_N^{\nu = 0}]^2 \to \lambda_N^{\nu = 0}.$
For the chiral condensate we then obtain the approximate expression
\be
\Sigma^{(2)} = \frac 1 {V\lambda_{\rm min}}.
 \ee
Using the Banks-Casher relation, 
the smallest nonzero eigenvalue is approximately given by
\be
\lambda_{\rm min} \approx \frac\pi{2\Sigma V},
\ee
resulting in $ \Sigma^{(2)} \approx \frac \pi 2 \Sigma^{(1)}$
This calculation can be made rigorous (see \cite{leonid-JV}) with the result
that both condensates become equal. Since we compare the ratio of two
partition functions, it is essential that they are normalized correctly.
This has been studied in a random matrix framework \cite{leonid-JV} confirming
the above results.

Also within a random matrix framework, it turns out that for 
$\mu \ne 0$ the two condensate become only equal 
after the partition functions have been correctly normalized
\cite{lehner}.

\subsection{Lesson 9: Test of Algorithms}

Because random matrix models are exactly solvable and show all essential
features due to nonhermiticity, they are an ideal tool to test 
algorithms for lattice QCD at nonzero chemical potential. 
We mention two examples: i) The density of states method
was analyzed and tested in a chiral random matrix model at nonzero
chemical potential \cite{ambjorn} and has been applied successfully 
to lattice QCD \cite{schmidt,ejiri}.
ii) The radius of convergence has been determined \cite{misha-2006} for
algorithms
that rely on Taylor expansion \cite{Allton,gupta}
or analytical continuation in \dl \mu \dr \cite{lombardo,deforcrand}.

\section{Conclusions}

We have shown that the phase of the fermion determinant dramatically affects
the physics of the QCD partition function. For example, the phase diagram of
QCD and $|{\rm QCD }|$ are completely different. This is our main motivation
for studying the behavior of the average phase factor and try to understand
its relation with the spectrum of the Dirac operator. In particular, this
has been  done in the microscopic domain  where QCD is
given by a chiral random matrix theory that can be solved analytically. In this
lecture we have discussed nine questions where random matrix theory has
contributed significantly to our understanding of QCD at $\mu \ne 0$. 
Perhaps the most important
contributions are the failure 
of the quenched approximation and the explanation of the mechanism that
results in a discontinuity of the chiral condensate when the quark mass
crosses the imaginary axis.

{\bf Acknowledgments:} 
We wish to thank the organizer's of CAQCD 2008 for an inspiring week in
Minneapolis as well as all of our colleagues who has contributed to the work
reported here. This work was
supported in part by U.S. DOE Grant No. DE-FAG-88ER40388. We thank the 
Institute for Nuclear Theory at the University of Washington for its
hospitality and the Department of Energy for partial support during the
completion of this work.

\end{document}